\documentstyle[prd,aps,graphicx]{revtex}
\begin{document}
\def \a {\alpha}
\def \b {\beta}
\def \g {\gamma}
\def \G {\Gamma}
\def \d {\delta}
\def \eps {\varepsilon}
\def \ep {\epsilon}
\def \e {\eta}
\def \f {\phi}
\def \ffi {\varphi}
\def \j {\iota}
\def \th {\theta}
\def \vth {\vartheta}
\def \k {\kappa}
\def \l {\lambda}
\def \m {\mu}
\def \n {\nu}
\def \x {\xi}
\def \p {\pi}
\def \r {\rho}
\def \s {\sigma}
\def \t {\tau}
\def \ps {\psi}
\def \o {\omega}
\def \z {\zeta}
\def \L {\pounds}
\def \vu {\tilde u}
\def \der {\partial }
\def \nn {\nonumber}
\def \rov {\equiv}
\def \A {{\cal A}}
\def \BB {{\cal B}}
\def \C {{\cal C}}
\def \D {{\cal D}}
\def \E {{\cal E}}
\def \F {{\cal F}}
\def \GG {{\cal G}}
\def \K {{\cal K}}
\def \N {{\cal N}}
\def \L {{\cal L}}
\def \cN {\bar\N}
\def \vN {\tilde\N}
\def \M {{\cal M}}
\def \vM {\tilde \M}
\def \I {{\cal I}}
\def \J {{\cal J}}
\def \R {{\cal R}}
\def \CC {{\rm C}}
\def \U  {{\cal U}}
\def \T  {{\cal T}}
\def \bfi {\bar \varphi}
\def \br {\bar \rho}
\def \bz {\bar z}
\def \bt {\bar t}
\def \pul {{{\scriptstyle{\frac{1}{2}}}}}
\def \sn {\sin \th}
\def \cs {\cos \th}
\def \tg {\tan \th}
\def \ctg {\cot \th}
\def \csec {\csc \th}
\def \dcsec {\csc^2 \th}
\def \msn {\sin^{-1} \th}
\def \ct {\bar t}
\def \dsn {\sin^2 \th}
\def \mdsn {\sin^{-2} \th}
\def \tsn {\sin^3 \th}
\def \csn {\sin^4 \th}
\def \dcs {\cos^2 \th}
\def \tcs {\cos^3 \th}
\def \csin {\sqrt{1-(wu)^2}}
\def \dctg {\cot^2 \th}
\def \chd  {\cosh 2\d}
\def  \shd  {\sinh 2\d}
\def \mm {\mbox{\quad }}
\def \mv {\mbox{\qquad }}
\def \msip {\rightarrow}
\def \vsip {\longrightarrow}
\def \lkz  {\bigl(}
\def \pkz  {\bigr)}
\def \lvkz {\Bigl(}
\def \pvkz {\Bigr)}
\def \lvvkz {\biggl(}
\def \pvvkz {\biggr)}
\def \lhz  {\bigl[}
\def \phz  {\bigr]}
\def \lvhz {\Bigl[}
\def \pvhz {\Bigr]}
\def \lvvhz {\biggl[}
\def \pvvhz {\biggr]}
\def \lsz   {\bigl\{ }
\def \psz   {\bigr\} }
\def \pvsz {\Bigl\} }
\def \lvsz {\Bigr\{ }
\def \lvvsz {\Biggl\{}
\def \pvvsz {\Biggr\}}
\def \BE {\begin{equation}}
\def \EE {\end{equation}}
\def \BDM {\begin{displaymath}}
\def \EDM {\end{displaymath}}
\def \BEAH {\begin{eqnarray*}}
\def \EEAH {\end{eqnarray*}}
\def \BEA {\begin{eqnarray}}
\def \EEA {\end{eqnarray}}
\def \BM {\begin{math}}
\def \EM {\end{math}}
\def \BDM {\begin{displaymath}}
\def \EDM {\end{displaymath}}
\def \mm {\mbox{\quad }}
\def \mv {\mbox{\qquad }}
\def \msip {\rightarrow}
\def \vsip {\longrightarrow}
\draft
\title{Co-accelerated particles in the C-metric}
\author{V. Pravda\footnote{E-mail: {\tt pravda@math.cas.cz} }, 
A. Pravdov\' a\footnote{E-mail: {\tt pravdova@math.cas.cz} } }
\address{Mathematical Institute, 
Academy of Sciences, \protect\\
\v Zitn\' a 25,
115 67 Prague 1, Czech Republic }
\date{\today}
\maketitle
\begin{abstract}
With appropriately chosen parameters,
the~C-metric represents two uniformly accelerated black holes moving
in the opposite directions on the axis of the~axial symmetry
(the~$z$-axis). The~acceleration is caused by nodal singularities
located on the~$z$-axis.

In the~present paper, geodesics in the~C-metric are examined. 
In general  there exist three types of timelike or null geodesics in the~C-metric: 
geodesics describing particles 1) falling under the~black
hole horizon; 2) crossing the~acceleration horizon; and 3) orbiting
around the~$z$-axis and co-accelerating with the~black holes.

Using an effective potential, it can be shown that there exist 
stable timelike geodesics of the~third type if the~product of the~parameters
of the~C-metric, $mA$, is smaller than a certain critical value.
Null geodesics of the~third type are always unstable.
Special timelike and null geodesics of the~third type are also
found in an analytical form.
\end{abstract}

\section{Introduction and Summary}

The~C-metric is a vacuum solution of the Einstein equations
of the~Petrov type D. Kinnersley and Walker \cite{KW}
showed that it represents black holes uniformly accelerated
by nodal singularities
in opposite directions along the~axis of the~axial symmetry. 
In coordinates \mbox{$\{ x$, $y$, $p$, $q\}$}
adapted to its algebraical
structure,  the~C-metric reads as follows:
\BE
{\rm d} s^2 = \frac{1}{A^2(x+y)^2} \left(
G^{-1}{\rm d}x^2+ F^{-1}{\rm d}y^2+
G {\rm d}p^2- F {\rm d}q^2  \right)  \ , \label{Cmetric}
\EE
where the~functions $F$, $G$ are the~cubic polynomials
\BEA
F&=&-1+y^2-2mAy^3 \ ,\\
G&=&\ \  1-x^2-2mAx^3 \ ,
\EEA
with  $m$ and $A$ being constant.
As we are choosing the~signature $+2$, we take $G>0$.

Although this form of the~C-metric (\ref{Cmetric}) is simple,
it is not suitable for physical interpretation of the~solution.
Since the~metric (\ref{Cmetric}) has two Killing vectors
$\der / \der p$, $\der /\der q$ it is possible to transform
it in its static regions given by
\BE
G>0\ ,\mm F>0 \label{static}
\EE
to the~Weyl form \cite{Bonnor} (see Sec.~\ref{secweyl}).
By further transformation \cite{Bonnor} one can show
that the~C-metric is in fact a radiative boost-rotation
symmetric spacetime \cite{Bonnor,AV} (see Sec.~\ref{secboost}).
The~class of boost-rotation symmetric spacetimes is the~only
class of exact radiative solutions of the~full nonlinear
Einstein equations that are known in an analytical form,
describe moving objects, and
are asymptotically flat (see \cite{BicSchPRD,JibiEhlers,AV}
for general treatise). Several generalizations of the~C-metric 
are known, let us mention the~charged C-metric \cite{KinnWalker,corn2}
and the~spinning C-metric \cite{PlebDem,bivoj}.

Farhoosh and Zimmermann \cite{FZ} 
studied a special class
of geodesics in the~C-metric -- test particles
moving on the~symmetry axis. 
We examine general geodesics starting in the~most
physical static region of the~C-metric (the~region
${\cal B}$, see Fig. \ref{regpq}) with the~help
of an effective potential.
It turns out that there exist three types of timelike  (null) geodesics: 
1) geodesics describing particles
falling under the~black hole horizon and then
on the~curvature singularity; 2) those ones describing
particles crossing the~acceleration horizon and
reaching future timelike (null) infinity -- they
are not co-accelerated with the~black holes; and 
3) geodesics describing particles spinning around 
the~axis of the~axial symmetry (the~$z$-axis),
co-accelerating with the~black holes along this axis
and reaching future null infinity. 
We investigate stability of timelike and null geodesics
of the~third type using the~effective potential
in the~coordinates \mbox{$\{ x$, $y$, $p$, $q\}$}
(Sect.~\ref{secxypq}) and in the~Weyl coordinates
(Sec.~\ref{secweyl}) in which it is easy to see
that the~stability of timelike geodesics does not
depend on the~distribution of conical singularities
located on the~$z$-axis.
We show that null geodesics of this type are always unstable
and that there exist stable timelike geodesics of the~considered
type if the~product of the~parameters of the~C-metric,
$mA$, is smaller than a certain critical value (\ref{numcond}).
This result indicates that  a black hole or a star, having satellites
in the~equatorial plane, which starts to accelerate in the~direction
perpendicular to this~equatorial plane can retain
some of its satellites  (which  is in fact very not
surprising) only if the~acceleration is sufficiently small.

We present special geodesics of the~third type in 
an analytical form representing particles 
(or zero-rest-mass particles) orbiting around 
the~axis of the~axial symmetry in a {\it constant distance}
and uniformly accelerating along the~$z$-axis 
(dragged by the~black holes).
They are given 
by $x=$ const, $y=$ const 
in the~coordinates \mbox{$\{ x$, $y$, $p$, $q\}$} (see Sec.~\ref{secxypq}),
by $\br =$ const, $\bz =$ const
in the~Weyl coordinates (Sec.~\ref{secweyl}) and 
finally we examine them in the~coordinates 
adapted to the~boost-rotation symmetry ($\r =$const, $z^2-t^2 =$ const)
where their physical interpretation can be easily understood (Sec.~\ref{secboost}).

In \cite{Cornish} it was shown that  the~Schwarzschild metric 
can be obtained from the~C-metric in the~Weyl coordinates
by the~limiting procedure $A\msip 0$ (note that one can 
get the~Weyl coordinates used in \cite{Cornish}
by multiplying the~Weyl coordinates in our paper by the~factor $A$).
Using this procedure one can find that the~unstable null geodesic
in the~C-metric representing photon-like particles orbiting
around the~$z$-axis corresponds to the~unstable circular 
photon orbit in the~Schwarzschild metric.




\section{Geodesics in  \lowercase{ $\{ x,\  y,\  p,\  q \}$} coordinates}
\label{secxypq}

\begin{figure}
\begin{center}
\includegraphics*[height=5cm]{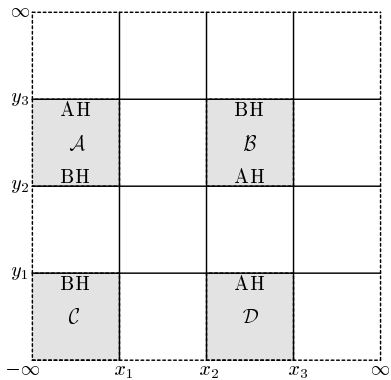}
\end{center}
\caption{The~character of the~regions determined by roots of the~polynomials $F$ and $G$ 
is schematically  illustrated.
Static regions are shaded, black hole and acceleration horizons
are denoted by BH and AH, respectively.
Dashed lines represent curvature singularities.}
\label{regpq}
\end{figure}

First we study geodesics in the~coordinates
$\{ x$, $y$, $p$, $q\}$ in which the~C-metric has the~form 
(\ref{Cmetric}). 
The~polynomials $F$ and $G$ entering the~metric
 have three different real roots iff the~condition
\BE
27 m^2 A^2 < 1 \label{exist}
\EE
holds. Then the~C-metric contains four different static regions 
${\cal A}, {\cal B}, {\cal C}, {\cal D} $ 
(see Fig. \ref{regpq}) where the~polynomials $F$, $G$ satisfy 
(\ref{static}). The~metric (\ref{Cmetric}) in each of these static
regions can be transformed to the~Weyl form with different
metric functions (see \cite{AV}). 

Curvature invariants diverge at  
$x \rightarrow \pm \infty$ or  $y \rightarrow \pm \infty$ 
 \cite{AV} where curvature singularities are located 
(see Fig.~\ref{regpq}).
The~region ${\cal B}$ is the~only static region without curvature
singularities. In fact, it is the~most physical static region
describing a uniformly accelerated black hole (see \cite{AV}
for analysis of the~other regions).
Acceleration horizons and black hole horizons are at $y=y_i$
where $y_i$ are the~roots of the~equation $F=0$ and are denoted
in Fig.~\ref{regpq} by AH and BH, respectively.

In this paper we consider only the~region ${\cal B}$ 
which exists iff the~condition (\ref{exist}) is satisfied. 

If a metric has a Killing vector $\x^\a$ then there exists a conserved
quantity $\x^\a U_\a$ for timelike geodesics 
with a tangent vector $U^\a$
and $\x^\a k_\a$ for null geodesics with a tangent vector $k^\a$.
Since the~C-metric (\ref{Cmetric}) has two Killing vectors
$\partial /\partial p$ and $\partial /\partial q$,
corresponding covariant  components of the 4-velocity,
$U_p$, $U_q$, for particles and components $k_p$, $k_q$
of the~wave 4-vector for zero-rest-mass particles
are conserved along geodesics and thus
\BEA
\frac{{\rm d}p(\tau)}{\rm d \tau}  &=& 
     L A^2 \frac{\lvkz x(\tau ) +y(\tau )\pvkz^2}
              {G(x(\tau ))} \ , 
\label{eqp} \\
\frac{{\rm d} q(\tau)}{\rm d \tau} &=& 
     E A^2 \frac{\lvkz x(\tau )+y(\tau )\pvkz^2}
               {F(y(\tau ))} \ ,
\label{eqq}
\EEA
where $L$ and $E$ are constants of motion, $\t$ is a proper time for
timelike geodesics and an affine parameter for null geodesics.

\begin{figure}
\begin{center}
\includegraphics*[height=4cm]{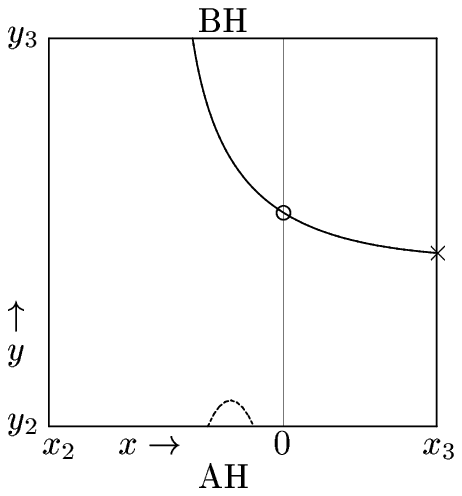} 
\hspace{2cm}
\includegraphics*[height=4cm]{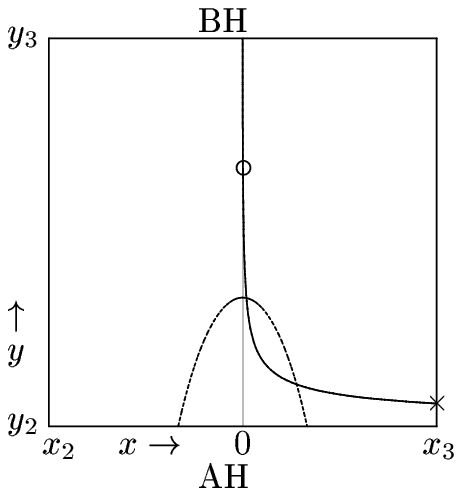} 
\end{center}
\caption{The~solid curve is given by (\ref{conditionxy})
(the~circle represents the~null geodesic given by (\ref{nullgeod}),
each point
on the~curve between the~circle and the~cross represents a timelike 
geodesic (\ref{timegeod_xy}) and the~remaining points correspond to
spacelike geodesics), 
the~dashed curve  is given by
$V,_{xx} V,_{yy}-{V,_{xy}}^2=0$, where $L$ was substituted from (\ref{resC1})
after performing the derivatives for:
a) $m=1/2$, $A=1/3$ (not satisfying (\ref{numcond}));
b) $m=0.02$, $A=0.05$ (satisfying (\ref{numcond})) -- 
between the~intersections of the~two plotted curves the~geodesics (\ref{timegeod_xy})
are stable and the~corresponding potential $V$ has there
its local minimum.}
\label{Figconditionxy}
\end{figure}

Let us examine special geodesics
$x(\tau)=x_0$, $y(\tau)=y_0$ with $x_0$, $y_0$ being constants.
Then substituting (\ref{eqp}), (\ref{eqq}) 
into the~geodesic equations we obtain
\BEA
{\frac {\left (1+mA{x_0}^{3}+x_0 y_0+3\,mA{x_0}^{2}y_0\right ){ L}^{2}}{{
\it G(x_0)}}}-{\frac {{\it G(x_0)}\,{ E}^{2}}{{\it F(y_0)}}}  &=& 0 \label{geod1} \ ,  \\
{\frac {{\it  F(y_0)}\,{ L}^{2}}{{\it G(x_0)}}}-{\frac {\left (-1+mA{y_0}^
{3}-x_0 y_0 +3\,mAx_0{y_0}^{2}\right ){ E}^{2}}{{\it F(y_0)}}} &=& 0 \label{geod2} \ .
\EEA 
A linear combination of these two equations leads to the~condition
\BE
3m^2A^2x_0^2y_0^2+mA\left (x_0 y_0+3\right )
\left (y_0-x_0\right )-1=0\ . \label{conditionxy}
\EE
Points $[x_0$, $y_0]$ in the~region ${\cal B}$
satisfying this condition are plotted in Fig.~\ref{Figconditionxy}.

The norm of the~four-velocity is
\BE
{A}^{2}\left (x_0+y_0 \right )^{2}\left ({\frac {{ L}^{2}}{{\it G(x_0)}}}
-{\frac {{ E}^{2}}{{\it F(y_0)}}}\right ) = \epsilon\ , \label{normcond}
\EE
where $\epsilon=-1$, $0$ for timelike and null geodesics, respectively. 

From (\ref{geod1}) and (\ref{normcond}) for a given timelike geodesic
($\epsilon=-1$), i.e. given $x_0$, $y_0$, 
constants $L$ and $E$ read
\BEA
L^2 &=& \frac{ G(x_0)^2}
 {A^2\left (x_0+y_0 \right)^3 \left( 1+ 3 mAx_0\right) x_0} 
\label{resC1} \ ,\\
E^2 &=& \frac{ F(y_0)^2}
 {A^2\left (x_0+y_0 \right)^3 \left( 1- 3 mAy_0\right) y_0} 
\ .\label{resC2}
\EEA

In the~region ${\cal B}$, $L^2$ and $E^2$ are positive only 
for $x_0>0$ where thus timelike geodesics exist 
(see Fig.~\ref{Figconditionxy}). Thus all timelike geodesics 
in the~region ${\cal B}$ 
with $x$, $y$ being constant are given by
\BEA
x(\t)&=&x_0\ ,\nn\\
y(\t)&=&y_0\  ,\nn\\
p(\t)&=&LA^2\frac{(x_0+y_0)^2}{G(x_0)}\ \t\ ,\label{timegeod_xy}\\
q(\t)&=&EA^2\frac{(x_0+y_0)^2}{F(y_0)}\ \t\ ,\nn
\EEA
where $x_0 \in (0$, $x_3)$, $y_0$ is given by (\ref{conditionxy}),
and $L$, $E$  are given by (\ref{resC1}) and (\ref{resC2}), respectively.

Spacelike geodesics have $x_0<0$ and we do not consider them further.

There exists a null geodesic ($\epsilon=0$, the~circle in Fig.~\ref{Figconditionxy}):
\BEA
x(\tau)&=&0\ , \nn \\
y(\tau)&=&\frac{1}{3m A} \  ,\nn\\
p(\t)&=& 
L\frac{1}{9m^2}\ \t\ ,\label{nullgeod}\\
q(\t)&=&E\frac{3A^2}{1-27m^2A^2}\ \t 
\ ,\nn\\
L^2& =& E^2\frac{27 m^2 A^2}{1-27 m^2 A^2} \ . \nn
\EEA
Notice that $L^2$ is positive since the~condition (\ref{exist}) is assumed
to be satisfied.

Let us now examine stability of {\it general} timelike geodesics. 
For this purpose we first construct an effective potential
of a general freely falling particle whose 4-velocity
has the~norm
\BE
-1=\frac{-1}{A^2(x+y)^2}\lvhz 
           -\frac{1}{G}\lvkz\frac{{\rm d}x}{{\rm d}\t}\pvkz^2
           -\frac{1}{F}\lvkz\frac{{\rm d}y}{{\rm d}\t}\pvkz^2
           -G\lvkz\frac{{\rm d}p}{{\rm d}\t}\pvkz^2
           +F\lvkz\frac{{\rm d}q}{{\rm d}\t}\pvkz^2\pvhz
\ .\label{tau}
\EE
Substituting from (\ref{eqp}), (\ref{eqq}) into (\ref{tau}) 
we get the condition 
\BE
\frac{F}{E^2A^4(x+y)^4}\lvvhz 
        \frac{1}{G}\lvkz \frac{{\rm d}x}{{\rm d}\t}\pvkz^2
       +\frac{1}{F}\lvkz \frac{{\rm d}y}{{\rm d}\t}\pvkz^2
        \pvvhz
    = \frac{E^2-V^2}{E^2}\ ,\label{podmE_V}
\EE
where the effective potential $V$ has the form
\BE
V=\sqrt{F\lvvkz \frac{1}{A^2(x+y)^2}+\frac{L^2}{G}\pvvkz}\ .\label{potent}
\EE
From (\ref{podmE_V}) it follows that a freely falling particle
with given $E$ has access only to those regions where $E>V$.
The~potential  $V$ goes to infinity for $x\msip x_2$ or $x\msip x_3$
and consequently particles cannot reach left and right edges of the~square ${\cal B}$.
Thus there remain only three possibilities: 1) particles leave the~square
${\cal B}$ across the~upper edge, i.e.,  they fall under the~black hole
horizon (on which $V=0$); 2) particles leave the~square ${\cal B}$ through
the~lower edge (on which also $V=0$), i.e., they cross the~acceleration
horizon; and 3) particles remain in the~square ${\cal B}$, i.e., they are
co-accelerated with the~black holes (see Sec.~\ref{secboost}).

Let us now study stability of geodesics of the~third type. There
exist stable geodesics of this type if $V$ has its local minimum
in the~region ${\cal B}$, i.e., if there exists a point ($x^*$, $y^*$)
in the~region ${\cal B}$ where
\BEA
&& V,_x (x^*,y^*)= V,_y (x^*,y^*)=0\ ,\label{VxVy}\\
&&V,_{xx}(x^*,y^*)V,_{yy}(x^*,y^*)-{V,_{xy}}^2(x^*,y^*)>0\ ,\label{VxxVyyVxy}\\
&&V,_{xx}(x^*,y^*)>0\ .\label{Vxx}
\EEA
For given $m$,  $A$, and $L$ there is only one point ($x^*$, $y^*$) 
in the~region ${\cal B}$ satisfying (\ref{VxVy}). It  lies 
on the curve (\ref{conditionxy}) with $L$ given by (\ref{resC1}),
i.e., it corresponds to the~geodesic (\ref{timegeod_xy}).
The~condition (\ref{Vxx}) is always satisfied, however, the~condition
(\ref{VxxVyyVxy}) is quite complicated and numerical calculations 
(see also Fig.~\ref{Figconditionxy} and the~text bellow) 
show that the~necessary condition for satisfying (\ref{VxxVyyVxy}) is
\BE 
mA < \sim 4.54 \times 10^{-3} \ .  \label{numcond} 
\EE

\begin{figure}
\begin{center}
\includegraphics*[height=5cm]{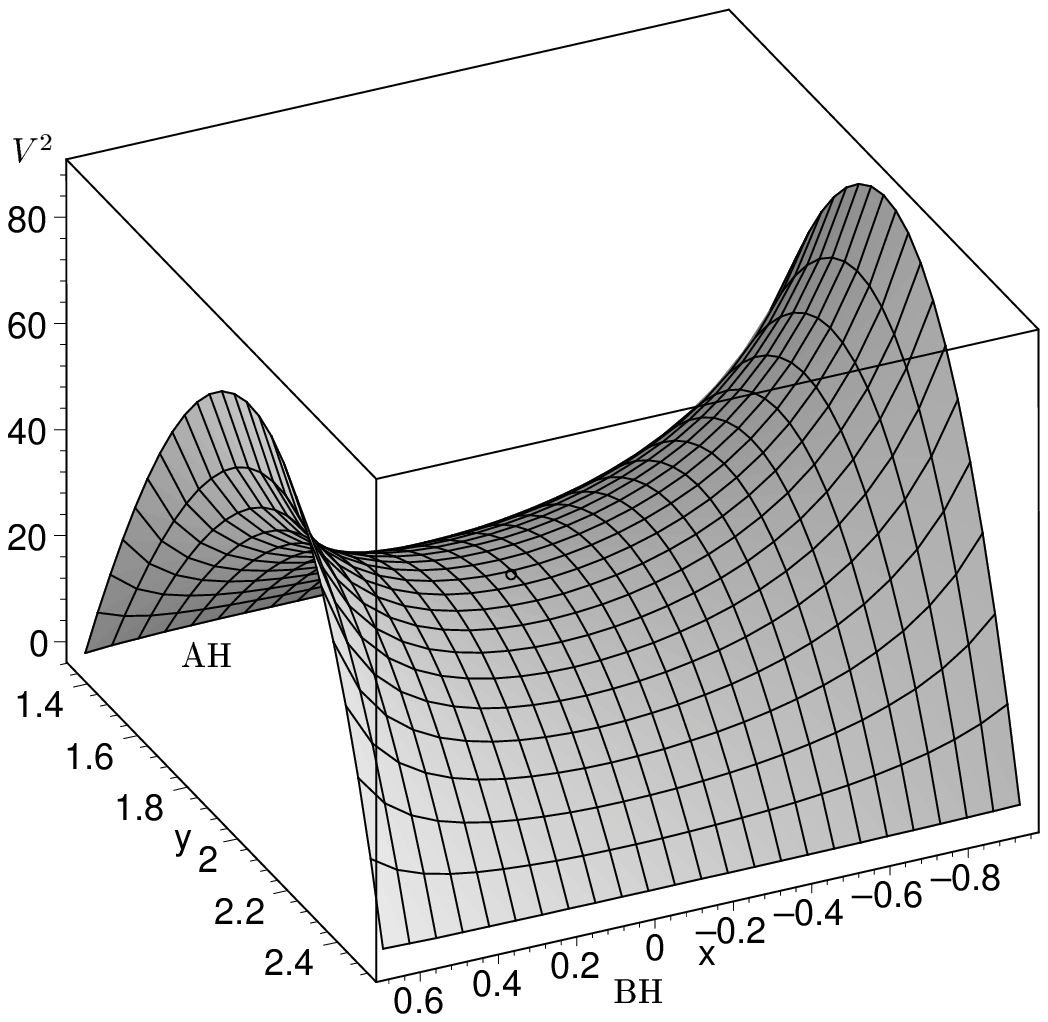}
\hspace{2cm}
\includegraphics*[height=5cm]{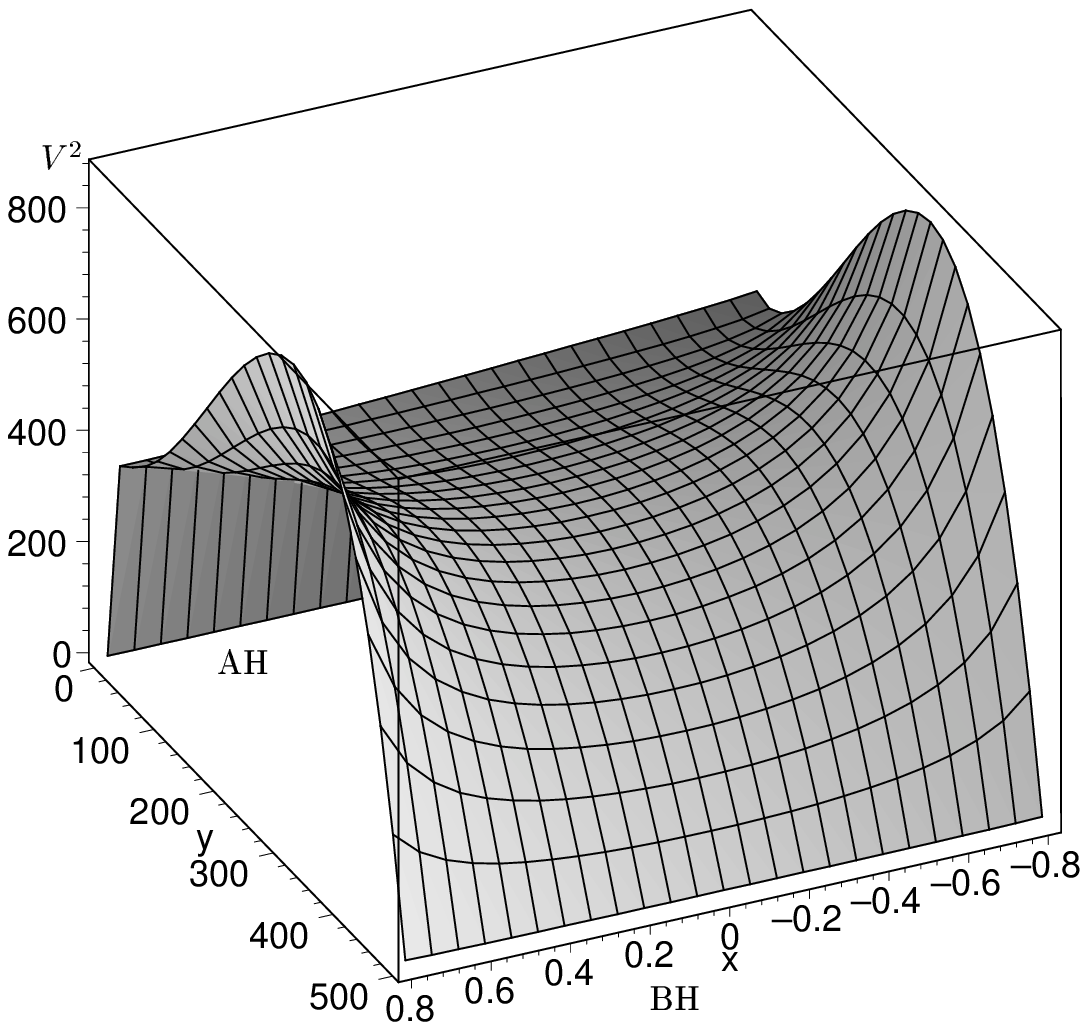}
\end{center}
\caption{The~function $V$ as a function of $x$, $y$ for $L=0.084$ and
a) $m=1/2$, $A=1/3$ (not satisfying (\ref{numcond})); 
b) $m=0.02$, $A=0.05$ (satisfying (\ref{numcond})), where
a local minimum exists.}
\label{potenc3d}
\end{figure}

Figs. \ref{potenc3d}a, b  illustrate the behaviour of the potential $V$ 
for parameters $m$, $A$ which do not satisfy and do satisfy 
the~condition (\ref{numcond}), respectively.

For parameters $m$, $A$ not satisfying the~condition (\ref{numcond}) 
there is no local minimum of the~potential $V$ and thus 
a small perturbation causes that a freely falling particle moving
along the~geodesic (\ref{timegeod_xy}) falls either under the black
hole horizon or under the~acceleration horizon.

\begin{figure}
\begin{center}
\includegraphics*[height=129pt]{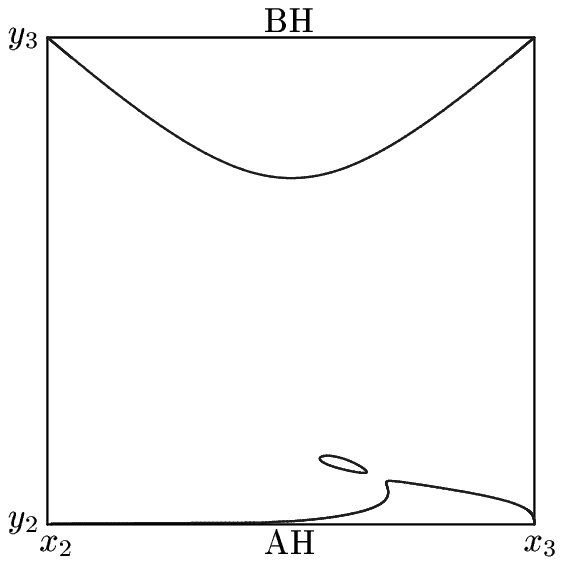}
\hspace{28pt}
\includegraphics*[height=129pt]{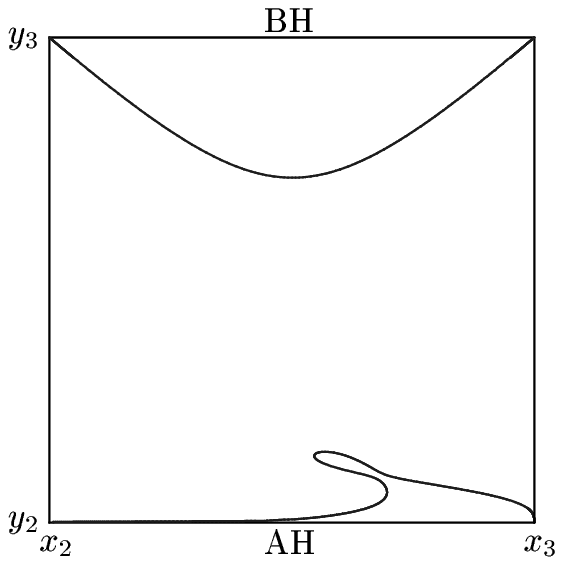}
\hspace{28pt}
\includegraphics*[height=129pt]{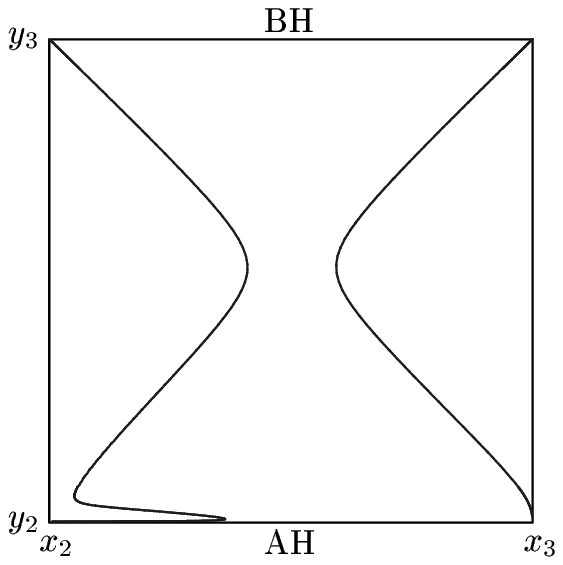}
\end{center}
\caption{Curves  $V$=const  for $m=0.02$, $A=0.05$, $L=0.084$:
a) $V=\sqrt{372.9}$;  b) $V=\sqrt{373}$; c) $V=\sqrt{429}$.}
\label{stability}
\end{figure}

\begin{figure}
\begin{center}
\includegraphics*[height=132pt]{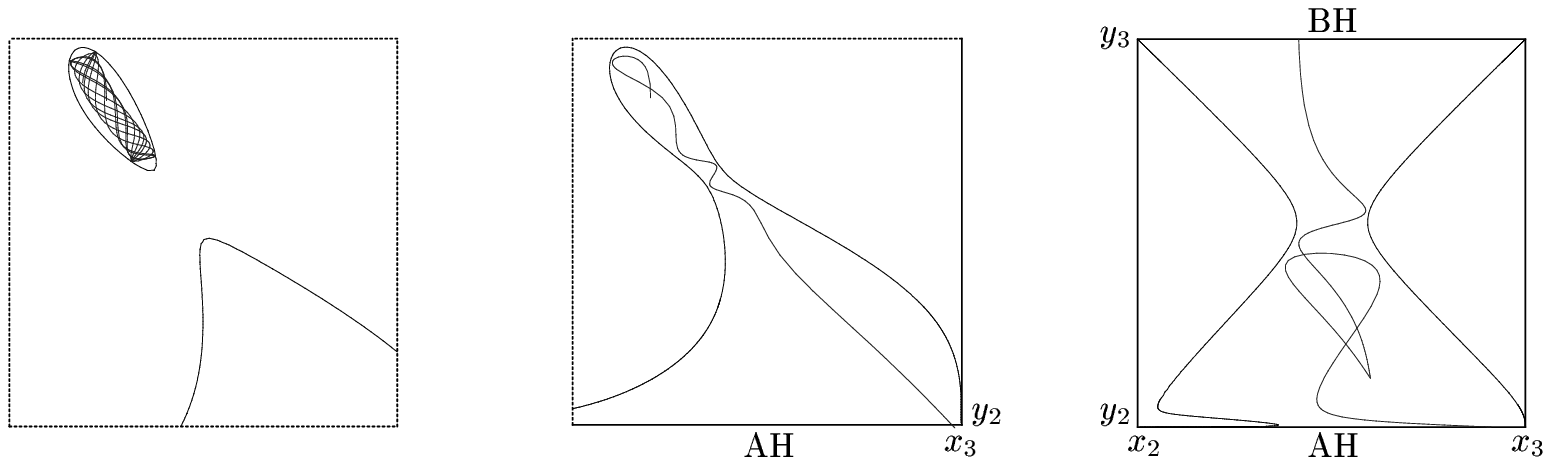}
\end{center}
\caption{Curves $V=$ const as in Fig.~\ref{stability}
and numerically obtained geodesics  for $m=0.02$, $A=0.05$, $L=0.084$.
All geodesics start at the~same point ($x_0$, $y_0$) which lies
on the~curve (\ref{conditionxy}) but have different initial velocities, 
i.e.  different constants of motion $E$:
a) $E=\sqrt{372.9}$ -- a geodesic of a co-accelerated particle;  
b) $E=\sqrt{373}$   -- a geodesic of a particle crossing the~acceleration
horizon; 
c) $E=\sqrt{429}$   -- geodesics of particles which fall under 
the~acceleration or the~black hole horizon depending on direction
of an initial velocity.}
\label{pic_stab_detail}
\end{figure}

For parameters $m$, $A$ satisfying the~condition (\ref{numcond})
and for suitable $L$ (see Fig.~\ref{Figconditionxy}) 
there exists a region (the~region bounded by a closed curve 
in Fig.~\ref{stability}a) from which freely falling particle with 
$E$ lower than a certain critical value cannot escape
(see Fig.~\ref{pic_stab_detail}a).
In this region considered geodesics are stable.
As it will be clear in Sec.~\ref{secboost} these trapped particles are
co-accelerated with the~uniformly accelerated black hole.
If the~parameter $E$ of these trapped particles is increased over a certain
critical value they fall under the~acceleration horizon 
(and thus they are not co-accelerated with the~black hole,
see Figs.~\ref{stability}b, \ref{stability}c, \ref{pic_stab_detail}b, 
\ref{pic_stab_detail}c) or under
the~black hole horizon (see Figs.~\ref{stability}c, 
\ref{pic_stab_detail}c).

Similarly we may derive an effective potential $\Lambda$
for zero-rest-mass  test particles substituting (\ref{eqp}), (\ref{eqq})
into the relation for the norm of the~wave 4-vector $k_{\a} k^{\a} = 0$:
\BE
\frac{F}{E^2A^4(x+y)^4}\lvvhz 
        \frac{1}{G}\lvkz \frac{{\rm d}x}{{\rm d}\t}\pvkz^2
       +\frac{1}{F}\lvkz \frac{{\rm d}y}{{\rm d}\t}\pvkz^2
        \pvvhz
    = 1- \frac{L^2}{E^2}\frac{1}{\Lambda^2}\ ,\label{nullpodmE_V}
\EE
where
\BE
\Lambda = \sqrt{\frac{G}{F}}\ .\label{Lambda}
\EE
Thus photon-like particles can reach only regions where
$L/E<\Lambda$.
In the~considered part ${\cal B}$ of the~spacetime,
the~function $\Lambda$ has vanishing first derivatives
at the~point \mbox{$x=0$}, \mbox{$y=1/(3mA)$}
(the~circle in Fig.~\ref{Figconditionxy}), however, 
there is not a local extreme and thus the~geodesic (\ref{nullgeod})
is unstable.


\section{Geodesics in the~Weyl coordinates}
\label{secweyl}


For interpreting the~geodesics (\ref{timegeod_xy}), (\ref{nullgeod})
we transform them into the~Weyl coordinates in this section and
into the~coordinates adapted to the~boost and rotation symmetries in the~next
section.

As was mentioned earlier the~C-metric in each of its static regions
can be transformed to the~static Weyl form
\BE
{\rm d}s^2={\rm e}^{-2U} 
\left[{\rm e}^{2 \nu}({\rm d}{\bar \rho}^2+{\rm d}{\bar z}^2)+
{\bar \rho}^2 {\rm d} \bar  \phi^2 \right] 
- {\rm e}^{2U} {\rm d} {\bar t}^2 \  \label{Weylmetric}
\EE
by transformation (see \cite{Bonnor})
\BEA
{\bar z}&=&\frac{1+mAxy(x-y)+xy}{A^2(x+y)^2} \ , \nn  \\
{\bar \rho}&=&\frac{\sqrt{FG}}{A^2(x+y)^2}\ , \label{transfWeyl} \\
{\bar \f}&=&p\ ,\nn\\
{\bar t}&=&q\ .\nn
\EEA

Transforming the~C-metric in the~region ${\cal B}$
into the~Weyl coordinates we obtain
\BEA
{\rm e}^{2U} &=&\frac{ \left[R_1-({\bar z}-{\bar z}_1) \right]  
      \left[R_3-(\bar z-{\bar z}_3) \right]}
{R_2-(\bar z-{\bar z}_2) }\ ,  \\
{\rm e}^{2\nu} &=& \frac{1}{4} 
 \frac{m^2}{A^6 ({\bar z}_2-{\bar z}_1)^2 ({\bar z}_3-{\bar z}_2)^2} 
\frac{\left[R_2 R_3 + {\bar \rho}^2 +({\bar z}-{\bar z}_2)({\bar z}-{\bar z}_3) \right]
      \left[R_1 R_2 + {\bar \rho}^2 +({\bar z}-{\bar z}_1)({\bar z}-{\bar z}_2) \right]}
{R_1 R_2 R_3 
\left[ R_1 R_3 + {\bar \rho}^2 
+ ({\bar z}-{\bar z}_1)({\bar z}-{\bar z}_3) \right]}\  {\rm e}^{2U}\ , 
\EEA
where functions $R_1$, $R_2$, and $R_3$ are defined by
\BE
R_i=\sqrt{({\bar z}-{\bar z}_i)^2+{\bar \rho}^2} \ ,\label{RiCmetric}
\EE
and  ${\bar z}_1 < {\bar z}_2 < {\bar z}_3$ are the~roots of the~equation
\BE
2A^4 {{\bar z}_i}^3 - A^2 {{\bar z}_i}^2 + m^2 =0 \ .
\EE

\begin{figure}
\begin{center}
\includegraphics*[height=4cm]{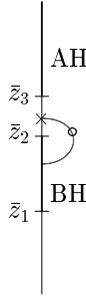}
\end{center}
\caption{The~region ${\cal B}$ in the~Weyl coordinates:
the~axis $\br=0$ with the~black hole horizon (BH, $\bz_1<\bz<\bz_2$)
and the~acceleration horizon (AH, $\bz>\bz_3$) and
the~curve given by Eq. (\ref{conditionR}) for $m=1/2$, $A=1/3$
(the~circle corresponds to the~null geodesic (\ref{geod_null_Weyl}), 
each point between the~circle and the~cross represents a timelike geodesic
(\ref{geodWeyl}) and the~other points on the~curve correspond to spacelike
geodesics).}   
\label{FigWeylaxis}
\end{figure}

As was shown in \cite{Bonnor}, the~C-metric in the~Weyl coordinates corresponds to the field
of  a rod between ${\bar z}_1$ and ${\bar z}_2$, 
a semi-infinite line mass at ${\bar z}>{\bar z}_3$ and
conical singularities for ${\bar z}<{\bar z}_1$ 
and ${\bar z}_2<{\bar z}<{\bar z}_3$ keeping them apart. 
The rod between ${\bar z}_1$ and ${\bar z}_2$ 
represents the black hole horizon and 
the~semi-infinite line mass at ${\bar z}>{\bar z}_3$ 
represents the acceleration horizon (see Fig.~\ref{FigWeylaxis}).

The~regularity condition of the~axis  
\BE
{\rm e}^{2\n}({\bar\r}=0,{\bar z} )=1\ \label{regulaxisWeyl}
\EE
is not satisfied at points where nodal singularities appear.
Since the~metric (\ref{Weylmetric}) with metric functions
\BEA
{\rm e}^{-2U'} &=& a\ {\rm e}^{-2U} \ , \label{newsol}\\
{\rm e}^{2\nu'} &=& b\ {\rm e}^{2\nu} \nonumber  \  ,
\EEA
$a$ and $b$ being constants, also satisfies the~vacuum Einstein
equations, by choosing the~constant $b$ appropriately
we may regularize (i.e. fullfil the~condition (\ref{regulaxisWeyl})) 
either the~part of the~axis 
${\bar z}<{\bar z}_1$ or 
${\bar z}_2<{\bar z}<{\bar z}_3$ (see \cite{AV}).

The~conditions (\ref{eqp}), (\ref{eqq}) for geodesics 
in the Weyl coordinates (corresponding to the~existence of two
Killing vectors $\der /\der{\bar t}$, $\der /\der{\bar\f}$)
read as follows
\BEA
\frac{\rm d {\bar \f}(\tau)}{\rm d \tau}  
    &=& L \frac{{\rm e}^{2U}}{ {\bar \r}(\tau)^{2}}\ , \label{eqphi} \\
\frac{\rm d{\bar t}(\tau)}{\rm d \tau} 
     &=& E {\rm e}^{-2U} \ .\label{eqt}
\EEA
Due to the~transformation (\ref{transfWeyl}), 
the~geodesics $x(\t)=x_0=$ const, $y(\t)=y_0=$ const, 
discussed in the previous section,  now have the~form
${\bar \r}(\t)=\br_0=$ const, ${\bar z}(\t)=\bz_0=$ const. 
The~condition (\ref{conditionxy}) in terms  of the~coordinates
 $\bar \rho$, $\bar z$ reads
\BE
R_1 R_3- R_3 R_2 - R_1 R_2 = 0 \ , \label{conditionR}
\EE
where $R_1$, $R_2$, $R_3$ are given by (\ref{RiCmetric})
and the~corresponding curve is plotted 
in Fig.~\ref{FigWeylaxis}.

Timelike geodesics (\ref{timegeod_xy}) 
have in the~Weyl coordinates the~form
\BEA
{\bar \r}(\t)&=&{\bar \r}_0\ ,\nn\\
{\bar z}(\t)&=&{\bar z}_0\ ,\label{geodWeyl}\\
{\bar \f}(\t)&=&L \frac{{\rm e}^{2U({\bar \r}_0,{\bar z}_0)} }
              {{\bar \r_0}^{2}}\t\ ,\nn\\
{\bar t}(\t)&=&E {\rm e}^{-2U({\bar \r}_0,{\bar z}_0)}\t\ ,\nn
\EEA
where ${\bar z}_0$, ${\bar \r}_0$ are constants satisfying 
(\ref{conditionR}), and  $E$, $L$
are constants of motion ((\ref{eqphi}), (\ref{eqt})).

Similarly the~null geodesic (\ref{nullgeod}) 
in the Weyl coordinates reads (the~circle in Fig.~\ref{FigWeylaxis})
\BEA
{\bar z}(\t)&=&9m^2\ ,\nn\\
{\bar \rho}(\t)&=& \sqrt{3(1-27 m^2 A^2)}\ \frac{m}{A}\ ,
\label{geod_null_Weyl}\\
{\bar \f}(\t)&=&L\frac{1}{9m^2}\ \t\ ,\nn\\
{\bar t}(\t)&=&E\frac{3A^2}{1-27m^2A^2}\ \t\ .\nn
\EEA


Analogously as in Sec.~\ref{secxypq}, (\ref{tau})--(\ref{potent}),
a motion of a freely falling particle 
with the~constants of motion $E$, $L$
is restricted to a region where $E>V$,
the effective potential $V$ being
\BE
V=\sqrt{{\rm e}^{2U}\lvkz 1+\frac{L^2 {\rm e}^{2U}}{{\bar \r}^2}\pvkz}\ .
\EE

Notice, that since the~potential $V$ and the~condition (\ref{conditionR}) 
do not depend on the~function ${\rm e}^{2\n}$,
after changing the~constant $b$, i.e. changing the~distribution of nodal
singularities, the~geodesics (\ref{geodWeyl})
remain geodesics and, moreover, this change does not affect their
stability.

\begin{figure}
\begin{center}
\includegraphics*[height=6cm]{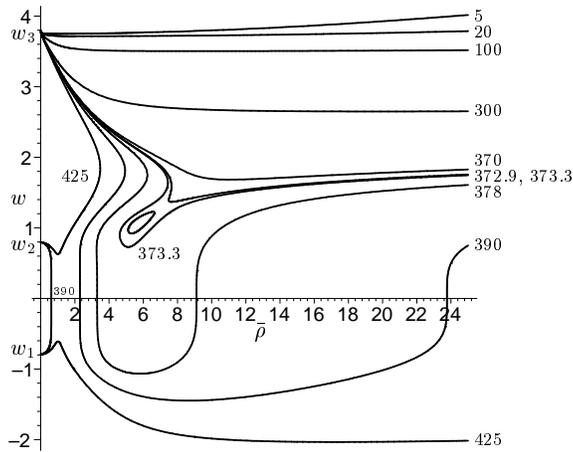}
\end{center}
\caption{Curves  $V^2$=const  for $m=0.02$, $A=0.05$, $L=0.084$ 
(to compactify the~picture coordinate 
$w=|{\bar z}|^{1/4} {\rm sign}\ {\bar z} $
is used instead of $\bz$; $w_i$ correspond to ${\bar z}_i$).
The~potential $V$ is infinite on the~axis at $w\in ( -\infty$, $w_1)$
and $w\in ( w_2$, $w_3)$ and null at the~black hole and
acceleration horizons at $w\in (w_1$, $w_2)$, $w\in (w_3$, $\infty)$,
respectively.}
\label{potencW}
\end{figure}

In Fig.~\ref{potencW} (analogous to Fig.~\ref{stability})
curves with different values of $V^2$ are plotted. There again appear
a region bounded by a closed curve from which trapped particles
with a given parameter $E$ cannot escape. Particles with $E$ higher
than a certain critical value fall either under the~black hole
horizon (the~axis between $w_1$, $w_2$ where 
$w=|{\bar z}|^{1/4} {\rm sig n}\ {\bar z}$)
or under the~acceleration horizon (the~axis between $w_3$ and $\infty$).

\section{Geodesics in the~canonical coordinates
adapted to the~boost-rotation symmetry}
\label{secboost}


To find an interpretation of geodesics studied in the~previous sections
we transform the~metric (\ref{Weylmetric}) by the~transformation 
\BEA
{\bar \rho}^2 &=& \rho^2 (z^2-t^2) \ ,\nn\\
\bar \phi &=& \phi\ , \label{trweylbs}\\
\bar z -{\bar z}_3 &=& \pul (t^2+\rho^2-z^2) \ ,\nonumber \\
\bar t &=& {\mbox{arctanh}} (t/z)\  \nonumber 
\EEA
into the form 
\BE
{\rm ds}^2 = -{\rm e}^{\lambda} {\rm d} \rho^2 - \rho^2 {\rm e}^{-\mu} {\rm d}\f^2 
- \frac{1}{z^2-t^2} \left[ ({\rm e}^{\lambda} z^2 - {\rm e}^{\mu} t^2 ) {\rm d} z^2
 -   2zt ({\rm e}^{\lambda} - {\rm e}^{\mu}  ) {\rm d} z\  {\rm d} t 
+     ({\rm e}^{\lambda} t^2 - {\rm e}^{\mu} z^2 ) {\rm d} t^2  \right]   
\label{BStvar}
\EE
which is adapted to the~boost and rotation symmetries (see \cite{BicSchPRD,AV}).
The inverse transformation to (\ref{trweylbs}) has the form
\BEA
\rho&=&\sqrt{\sqrt{\br^2+(\bz-\bz_3)^2}+(\bz-\bz_3)} \ , \nn \\
z&=&\pm\sqrt{\sqrt{\br^2+(\bz-\bz_3)^2}-(\bz-\bz_3)}\cosh \bt \ , \label{transfT1i}\\
t&=&\pm\sqrt{\sqrt{\br^2+(\bz-\bz_3)^2}-(\bz-\bz_3)}\sinh \bt  \ , \nonumber 
\EEA
where either upper or lower signs are valid.

\begin{figure}
\begin{center}
\includegraphics*[height=4cm]{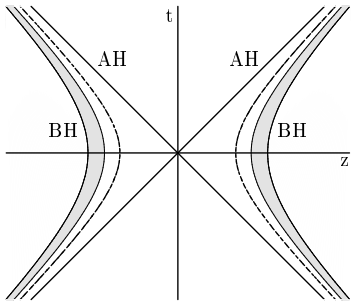}
\end{center}
\caption{Uniformly accelerated black holes (the~shaded region)
and co-accelerated test particles (dashed lines).}
\label{fig_boost}
\end{figure}

From (\ref{transfT1i}) it follows that 
geodesics we are interested in, satisfying
$\br=\br_0=$ const, $\bz=\bz_0=$ const in the~Weyl coordinates,
in the~coordinates $\{ t$, $\r$, $z$, $\f\}$
satisfy $\rho =$ const and $z^2-t^2=$ const 
(the~worldline is a hyperbola in the~($z$,$t$)-plane which
corresponds to a uniformly accelerated motion along the~$z$-axis,
see Fig.~\ref{fig_boost}).  
Geodesics of this type
(corresponding to (\ref{geodWeyl}), (\ref{geod_null_Weyl})) 
now have the form 
\BEA
\rho (\t)&=& K_1\ , \nonumber \\
\phi  (\t)&=& c_2 \tau \ ,\\
z      (\t)&=&\pm K_2 \cosh c_1 \tau\ , \nn\\
t      (\t)&=& K_2 \sinh c_1 \tau \  ,\nonumber 
\EEA
where the~constants $K_1$, $K_2$, $c_1$, and $c_2$  read
\BEA
K_1&=&\sqrt{\sqrt{\br_0^2+(\bz_0-\bz_3)^2}+(\bz_0-\bz_3)}\  ,\nn\\
K_2&=&\sqrt{z^2-t^2}=\sqrt{\sqrt{\br_0^2+(\bz_0-\bz_3)^2}-(\bz_0-\bz_3)}\  ,\nn\\
c_1&=&E{\rm e}^{-2U({\bar \r}_0,\ {\bar z}_0)}\ ,\nn\\
c_2&=&L \frac{{\rm e}^{2U({\bar \r}_0,\ {\bar z}_0)}}{{\bar \r_0}^{2}}\ .\nn
\EEA
These geodesics describe particles orbiting the~$z-$axis 
and uniformly accelerating
along the~$z$-axis. 

%
%
%
 
\end{document}